\documentclass[a4paper]{article}

\usepackage{INTERSPEECH2021}

\usepackage{subcaption} 
\usepackage{booktabs} 
\usepackage{multirow} 
\usepackage{ragged2e}
\usepackage{amssymb, pifont} 
\usepackage{colortbl}
\usepackage{subcaption}
\usepackage[dvipsnames]{xcolor}
\usepackage{pgfplots}

\pgfplotsset{compat=newest}
\usepgfplotslibrary{groupplots}

\newcommand{\B}{\textbf}
\newcommand{\U}{\underline}

\title{S2VC: A Framework for Any-to-Any Voice Conversion with Self-Supervised Pretrained Representations}
\name{
    Jheng-hao Lin \quad
    Yist Y. Lin \quad
    Chung-Ming Chien \quad
    Hung-yi Lee \quad
}
    
\address{College of Electrical Engineering and Computer Science, National Taiwan University, Taiwan}
\email{\{r08922049, r08922048, r08922080, hungyilee\}@ntu.edu.tw}

\begin{document}

\maketitle

\begin{abstract}
    Any-to-any voice conversion (VC) aims to convert the timbre of utterances from and to any speakers seen or unseen during training.
    Various any-to-any VC approaches have been proposed like \textsc{AutoVC}, AdaINVC, and FragmentVC.
    \textsc{AutoVC}, and AdaINVC utilize source and target encoders to disentangle the content and speaker information of the features.
    FragmentVC utilizes two encoders to encode source and target information and adopts cross attention to align the source and target features with similar phonetic content.
    Moreover, pretrained features are adopted. \textsc{AutoVC} used d-vector to extract speaker information, and self-supervised learning (SSL) features like wav2vec 2.0 is used in FragmentVC to extract the phonetic content information.
    Different from previous works, we proposed S2VC that utilizes Self-Supervised features as both source and target features for the VC model.
    Supervised phoneme posteriorgram (PPG), which is believed to be speaker-independent and widely used in VC to extract content information, is chosen as a strong baseline for SSL features.
    The objective evaluation and subjective evaluation both show models taking SSL feature CPC as both source and target features outperforms that taking PPG as source feature, suggesting that SSL features have great potential in improving VC.

\end{abstract}
\footnote{We acknowledge the support of AWS Machine Learning Research Awards program.}

\section{Introduction}
\label{sec:introduction}

Self-supervised learning (SSL)~\cite{ss_cv, bert} has obtained impressive results these years in different domains, including computer vision, natural language processing, and speech processing.
The self-supervised training regime does not rely on human annotations of the data, which is expensive to collect and thus benefits from the use of a large amount of unlabeled data.
SSL models pretrained on speech corpora have been shown to be able to extract speech representations that can be used in downstream tasks such as automatic speech recognition, speaker recognition, and speech translation~\cite{apc2, cpc, mockingjay}.

VC aims to convert a source utterance to sound like spoken by a target speaker while preserving the original phonetic content.
The conversion can be achieved by disentangling the content and speaker information from the source and target utterances, respectively, then combining them and synthesizing the converted utterance.
Supervised pretrained representations have long been used to provide content or speaker information for VC tasks.
Phoneme posteriorgram (PPG) especially is very popular among VC implementations~\cite{PPG_m2o_vc, PPG_a2m_vc, vcc2020}.
PPG is speaker-independent and suitable for removing the speaker characteristics from the voice to be converted.
The speaker representations pretrained by speaker recognition tasks such as d-vector~\cite{dvector} and x-vector~\cite{xvector} have been widely used to provide speaker information. 

On the other hand, SSL representations can be utilized for phoneme recognition and speaker classification~\cite{apc2, cpc, mockingjay}, which indicates that both phonetic and speaker information are hereditary contained in SSL representations and thus make them possible to be used in the voice conversion (VC) task.
Several previous works have tried to introduce SSL representations to the VC task.
For example, Huang et al.~\cite{vqwav2vec_a2o_vc} proposed a sequence-to-sequence VC framework where SSL representations were used to capture the phonetic information of the source utterance.
However, their approach can only convert the utterance to a predefined target speaker.
On the other hand, FragmentVC~\cite{fragmentvc} is an any-to-any VC model which can convert the speech of an arbitrary source speaker to any target speaker, even speakers unseen during the training time.
It also used pretrained SSL models to extract the content information from the source utterances.

In this paper, we aim to improve any-to-any VC (also called one-shot VC)~\cite{a2a_vc, adain_vc, autovc, VQVC+}, which is one of the most difficult VC settings, by involving various pretrained SSL representations.
Different from previous works~\cite{vqwav2vec_a2o_vc, fragmentvc}, we extract not only the phonetic information but also the target speaker information from the SSL representations.
Several different SSL models, including Autoregressive Predictive Coding (APC) ~\cite{apc2}, Contrastive Predictive Coding (CPC)~\cite{cpc_audio}, and wav2vec 2.0~\cite{wav2vec2}, are investigated, and we also compare the performance of these SSL representations with supervised representations such as PPG representation~\footnote{We did not compare with the supervised speaker representations because the recent work~\cite{vc_survey} showed that they are not suitable for delivering speaker information needed in VC.}.
The results show that SSL representations achieve comparable and even better performance than PPG representation on both subjective and objective evaluation.

\section{Methods}
\label{sec:methods}

\begin{figure*}[htb]
    \centering
    \vspace{-2.3em}
    \includegraphics[width=\linewidth]{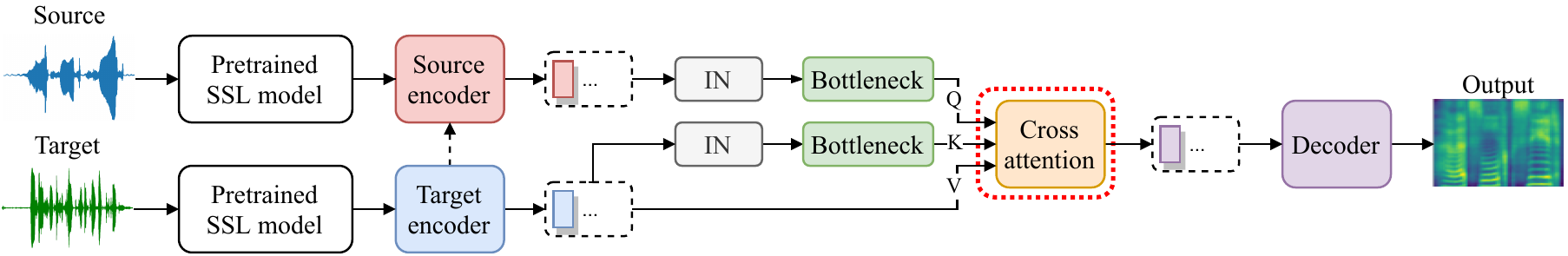}
    \caption{
        The overall model architecture of S2VC.
        IN denotes the Instance Normalization.
    }
    \label{fig:model_arch}
    \vspace{-1.5em}
\end{figure*}

The overall framework of S2VC is in Fig.~\ref{fig:model_arch}.
As shown in the figure, we adopt pretrained SSL models to extract source and target features.
In the following subsections, we first introduce the foundation of this work: FragmentVC~\cite{fragmentvc}, an architecture potential to utilize any kinds of speech representations.
Followed by the modification we made to FragmentVC to further improve the performance.
Then briefly describe several SSL features tested in our framework.

\begin{figure}[htb]
    \centering
    
    \vspace{-1em}
    \includegraphics[width=\linewidth]{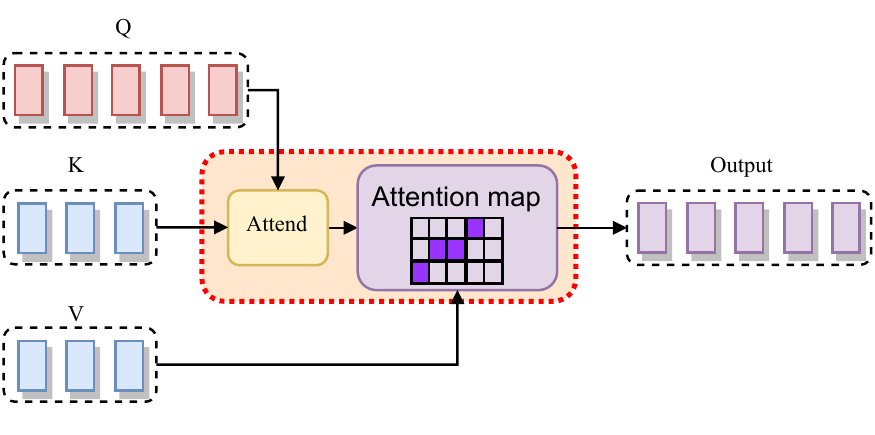}
    \caption{
        Conceptual illustration of cross attention.
    }
    \label{fig:cross_attn}
    \vspace{-1em}
\end{figure}

\subsection{Baseline: FragmentVC}
The overall framework evolves from FragmentVC.
FragmentVC is a deep exemplar-based model consisting of a source encoder, a target encoder, cross attention modules, and a decoder.
The conceptual illustration of cross attention is in Fig.~\ref{fig:cross_attn}, which takes one output feature from the source encoder (Q) and two output features from the target encoder (K, V). 
The output feature sequence of the target encoder (K) is then attended by the source encoder's output (Q).
In this architecture, it is found that the cross attention module learns to align the source features to the target features with similar phonetic content, which is similar to the idea of exemplar-based VC models.
The decoder then produces the converted Mel spectrogram from the attention-augmented features (V).
At training time, the same utterance is used as the input of both the source and the target encoder and the decoder's reconstruction target.
The encoders automatically learn to disentangle the content and the speaker information without any explicit constraint.

\subsection{Modifications}
We try to help the cross attention module to align the source and target features from the following two perspectives.
Self-attention pooling guides the representation encoded by the source encoder to be close to that encoded by the target encoder.
Attention information bottleneck removes redundant information in both representations encoded in Q and K to make the attention only consider the phonetic content information.

\subsubsection{Self-attention pooling}

Safar et al.~\cite{sap} proposed a self-attention pooling layer and showed it is good at extracting the time-invariant features information.
We utilize the self-attention pooling layer to extract the representation from the target encoder.
The extracted representation is then applied to the source encoder (dotted line in Fig.~\ref{fig:model_arch}) to guide the representation encoded by the source encoder closer to that by the target encoder.

\subsubsection{Attention information bottleneck}

AdaINVC~\cite{adain_vc} showed that Instance Normalization is capable of removing speaker-dependent information from the features, and AutoVC~\cite{autovc} used a carefully designed hidden dimension of the encoder layers to extract speaker-independent content information.
We combined them to the attention layer by applying Instance Normalization to both the Q and K, followed by a bottleneck layer so as to lower the speaker information encoded in them.

\subsection{SSL representations}

Three well-known SSL representations studied in this work are APC~\cite{apc2, apc1}, CPC~\cite{cpc_audio, cpc} and wav2vec 2.0~\cite{wav2vec2}.
APC learns the representation in a way similar to a conventional RNN-based language model.
It takes Mel spectrograms as input, and by predicting the future input conditioning on the past inputs, APC learns general speech representations in an auto-regressive way.
On the other hand, CPC and wav2vec 2.0 directly utilize waveforms as input.
CPC also learns the representation in an auto-regressive way but the prediction is done in the compact latent space instead of the input feature space, and the training is to optimize a probabilistic contrastive loss.
Wav2vec 2.0 further improves on CPC by enlarging the model size and replacing auto-regressive prediction with masked language model-like prediction similar to BERT.

In the original Fragment VC~\cite{fragmentvc}, the source encoder takes wav2vec 2.0 representations as inputs to extract a feature sequence containing content information of the source utterance, while the target encoder takes Mel spectrogram of several utterances of the target speaker as inputs.
Though only two representations are studied in FragmentVC, we consider this architecture and training scheme to be very flexible to incorporate any kinds of speech representations. 
This paper takes the representation of APC, CPC, or wav2vec 2.0 as the input features of source and target encoders and explores all the combinations.

\section{Experimental setup}
\label{sec:experimental_setup}

\subsection{Training Setup}

We trained all the models on CSTR VCTK Dataset~\cite{vctk} with 44 hours of audios spoken by 109 native speakers. All the audios were resampled from 48k Hz to 16k Hz before extracting the features.
The optimizer we used is AdamW~\cite{adamw} with learning rate 5e-5, and $\beta_1$ = 0.9, $\beta_2$ = 0.999.
The source encoder is composed of 4 linear layers with batch normalization.
The target encoder consists of 3 one-dimensional convolution layers.
The decoder is comprised of 3 conformer~\cite{conformer} layers followed by a linear projection to the Mel spectrogram's dimension.
The dimension of queries and keys in the cross attention is 4 for the models with bottleneck, and their dimension is 512 for those without bottleneck.
The L1 loss between the predicted and the ground-truth log Mel spectrogram is used.
Universal neural vocoding~\cite{universal_vocoder} trained on a combination of LJ-speech~\cite{ljspeech}, LibriTTS-train-clean-100, and CMU Arctic dataset~\cite{cmu} for 200k steps with batch size 32 is adopted as the vocoder of all the models.

\subsection{Feature Extraction}

We extract both PPG and SSL representations via S3PRL\footnote{https://github.com/s3prl/s3prl}~\cite{S3PRL} speech toolkit, which provides a user-friendly interface to extract various pretrained representations.
The PPG is pretrained on the TIMIT dataset~\cite{timit}, achieving 21.7\% frame-wise phone error rate (much more challenging measurement than phone error rate) on the test set.
As for the self-supervised features such as APC and CPC, the official pretrained checkpoints are used.

\subsection{Testing scenarios}

We evaluate the voice conversion models in the following two scenarios.
The first one is the conversion between speakers in the training dataset VCTK (s2s).
The other one is the conversion between speakers in the unseen dataset CMU (u2u).
For each scenario, we randomly sampled 400 testing pairs.
Each testing pair contains one utterance from the source speaker and five utterances from the target speaker.

\subsection{Objective evaluation}

We adopt two automatic assessment systems to evaluate the quality and the speaker similarity of converted utterances.
MOSNet~\cite{MOSNet} is adopted to efficiently assess the quality of the synthetic utterances.
Similar to the Mean Opinion Score (MOS) scored by human subjects, MOSNet takes an utterance as input and outputs a score ranging from 1.0 to 5.0, where the higher the score is, the better the quality is.

A publicly available pretrained speaker verification (SV) system\footnote{https://github.com/resemble-ai/Resemblyzer} is adopted to assess the speaker similarity between a converted utterance and a target utterance, as done in previous work~\cite{jia2018transfer}.
The SV system first extracts the utterance-level embeddings of the converted utterance and the target utterance, and the cosine similarity between the embeddings is computed as the similarity score.
The SV system accepts a converted utterance if the similarity score is higher than a pre-defined threshold which is computed by finding the EER throughout the dataset.
The percentage of converted utterances accepted by the SV system, which we call SV accuracy, is used as an evaluation metric.

\subsection{Subjective Evaluation}
To evaluate the perceptual quality of converted utterances, we conducted Mean Opinion Score (MOS) tests in quality and speaker similarity.
For the quality MOS test, the subjects listen to an authentic vocoder-reconstructed utterance or a converted utterance. Then they score it from 1.0 to 5.0 in terms of quality (1.0 means bad, and 5.0 means perfect).
For the similarity MOS test, the subjects listen to an authentic target utterance and a converted utterance. Then they score it from 1.0 to 5.0 in terms of speaker similarity (1.0 means very different and 5.0 means absolutely the same)~\cite{jia2018transfer}.
For every model considered, we evaluate the same 40 pairs of real and converted utterances, sampled from the 400 testing pairs, with each scored by at least 5 subjects.
The scores are reported with the 95\% confidence intervals for each model.
Since the u2u scenario is more challenging than that of s2s, the subjective evaluation is conducted for u2u.

\section{Results and analysis}
\label{sec:results_and_analysis}

\begin{table*}[htb]
    \setlength{\tabcolsep}{2pt} 
    \vspace{-2.5em}
    \caption{The overall objective results of MOSNet prediction and Speaker Verification}
    \label{tab:total}
    \centering
    \begin{subtable}{.49\linewidth}
        \centering
        \caption{
            The objective results of the MOSNet predictions on both s2s and u2u scenarios.
        }
        \label{tab:mosnet}
        \footnotesize
        \begin{tabular}{cccccc}
            \toprule
            \multirow{3}{*}{\B{Target}}
                        &
            \multicolumn{5}{c}{\B{Source}}
            \\
            \cmidrule(lr){2-6}
                        & Mel         & PPG         & APC         & CPC                 & W2V         \\
            \midrule
            Mel         & 3.36 / 2.74 & 3.00 / 2.84 & 3.05 / 2.94 & 3.25 / 2.81         & 3.44 / 2.93 \\
            PPG         & 2.94 / 2.90 & 2.46 / 2.82 & 3.06 / 3.09 & 3.30 / 3.19         & 2.95 / 2.80 \\
            APC         & 3.07 / 3.01 & 3.06 / 2.87 & 2.97 / 2.83 & 3.41 / 3.25         & 3.03 / 2.83 \\
            CPC         & 3.29 / 2.96 & 3.16 / 3.19 & 2.83 / 3.14 & 3.36 / 3.07         & 3.08 / 2.68 \\
            W2V         & 2.95 / 2.55 & 2.99 / 3.04 & 3.12 / 3.10 & 3.30 / 3.05         & 2.93 / 2.63 \\
            \midrule
            \B{Average} & 3.12 / 2.83 & 2.93 / 2.95 & 3.01 / 3.02 & \B{3.32} / \B{3.07} & 3.09 / 2.77 \\
            \bottomrule
        \end{tabular}
    \end{subtable}%
    \hfill
    \begin{subtable}{.49\linewidth}
        \centering
        \caption{
            The objective results of the speaker verification accept rate (\%) on both s2s and u2u scenarios.
        }
        \label{tab:sv}
        \footnotesize
        \begin{tabular}{cccccc}
            \toprule
            \multirow{3}{*}{\B{Source}}
                        &
            \multicolumn{5}{c}{\B{Target}}
            \\
            \cmidrule(lr){2-6}
                        & Mel                 & PPG         & APC         & CPC                 & W2V         \\
            \midrule
            Mel         & 93.0 / 97.3         & 12.3 / 20.8 & 69.0 / 92.3 & \U{98.8} / \U{98.0} & 14.3 / 5.8  \\
            PPG         & 78.8 / 95.3         & 14.5 / 20.3 & 81.3 / 96.0 & 78.8 / 88.3         & 69.0 / 73.8 \\
            APC         & 69.5 / 89.0         & 11.8 / 21.8 & 66.5 / 92.3 & 53.5 / 81.5         & 19.3 / 28.5 \\
            CPC         & 98.8 / 99.5         & 19.3 / 10.8 & 90.0 / 59.0 & \U{97.8} / \U{96.8} & 68.0 / 35.3 \\
            W2V         & 96.3 / 99.5         & 6.3 / 4.8   & 38.3 / 11.5 & 41.3 / 24.8         & 31.5 / 7.0  \\
            \midrule
            \B{Average} & \B{87.3} / \B{96.1} & 12.8 / 15.7 & 69.0 / 70.2 & 74.0 / 77.9         & 40.4 / 32.1 \\
            \bottomrule
        \end{tabular}
    \end{subtable}
    \justify{
        \centering
        \emph{A / B}: 'A' and 'B' are the result for s2s and u2u respectively.
        \emph{W2V}: wav2vec 2.0.\\
    }
    \vspace{-1.2em}
\end{table*}

\subsection{Objective performance analysis}

Five different representations, including Mel spectrogram, PPG, APC, CPC, and wav2vec 2.0, are used as the source or target feature of the VC models.

The results of MOSNet predictions are listed in Table~\ref{tab:mosnet}.
In this table, we compare the average MOSNet predictions of the models with different representations as source features to see the ability of the representations in providing content information.
The results show that both APC and CPC outperform PPG, suggesting that they may be promising to provide content information for VC.
The performances of Mel spectrogram and wav2vec 2.0 are not ideal in the u2u scenario, showing that they are relatively not robust to unseen data.

The results of SV accuracies are listed in Table~\ref{tab:sv}.
Here we compare the average SV accuracies of the models with different representations as target features to examine their capability in terms of providing speaker information.
The results show that Mel spectrogram is the best choice for extracting speaker-dependent information; however, some models with CPC (with underline in table) achieve comparable or better performance than those with Mel spectrogram, showing that CPC is also good at providing speaker-dependent information for VC.

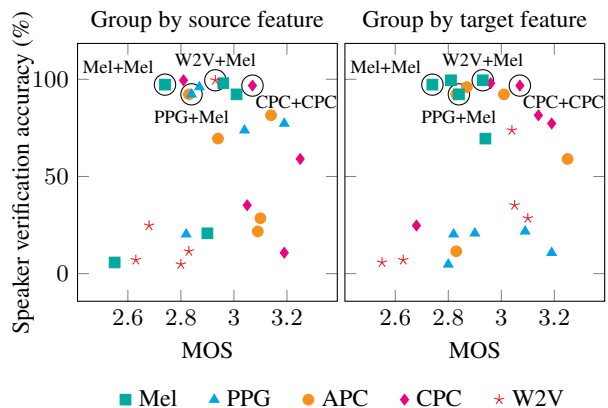
\begin{figure}[htb]
    \centering
    \pgfplotsset{
        every axis title/.append style={
                yshift=-5pt,
            },
        legend style={
                at={(0.0,-0.3)},
                anchor=north,
                legend columns=-1,
                draw=none,
                column sep=1pt,
                /tikz/every even column/.append style={column sep=10pt}
            },
    }
    \begin{tikzpicture}
        \begin{groupplot}[
                group style={
                        group size=2 by 1,
                        horizontal sep=0.1cm,
                    },
            ]
            \nextgroupplot[
                height=5.0cm,
                width=5.0cm,
                enlargelimits=0.2,
                xlabel={MOS},
                ylabel={Speaker verification accuracy (\%)},
                y label style={
                        yshift=-7pt,
                    },
                title={Group by source feature},
                scatter/classes={%
                        mel={mark=square*,Emerald},
                        ppg={mark=triangle*,Cerulean},
                        apc={mark=*,BurntOrange},
                        cpc={mark=diamond*,RubineRed},
                        w2v={mark=star,draw=Red}
                    },
            ]
            \addplot[
                scatter,mark=*,only marks,
                scatter src=explicit symbolic,
            ]
            table[
                    x=mos,
                    y=sv,
                    meta=source,
                ]{
                    mos   sv     source  target
                    2.74  97.25  mel     mel
                    2.90  20.75  mel     ppg
                    3.01  92.25  mel     apc
                    2.96  98.00  mel     cpc
                    2.55  5.75   mel     w2v
                    2.94  69.50  apc     mel
                    3.09  21.75  apc     ppg
                    2.83  92.25  apc     apc
                    3.14  81.50  apc     cpc
                    3.10  28.50  apc     w2v
                    2.84  92.25  ppg     mel
                    2.82  20.25  ppg     ppg
                    2.87  96.00  ppg     apc
                    3.19  77.25  ppg     cpc
                    3.04  73.75  ppg     w2v
                    2.81  99.50  cpc     mel
                    3.19  10.75  cpc     ppg
                    3.25  59.00  cpc     apc
                    3.07  96.75  cpc     cpc
                    3.05  35.20  cpc     w2v
                    2.93  99.50  w2v     mel
                    2.80  4.75   w2v     ppg
                    2.83  11.50  w2v     apc
                    2.68  24.75  w2v     cpc
                    2.63  7.00   w2v     w2v
                };
            \draw[black] (3.07, 96.75) circle (4.0pt) node[below right, xshift=-2pt, yshift=-0.5pt] {\scriptsize{CPC+CPC}};
            \draw[black] (2.74, 97.25) circle (4.0pt) node[above left, xshift=-1pt, yshift=1pt] {\scriptsize{Mel+Mel}};
            \draw[black] (2.84, 92.25) circle (4.0pt) node[below, yshift=-3pt] {\scriptsize{PPG+Mel}};
            \draw[black] (2.93, 99.50) circle (4.0pt) node[above, yshift=2pt] {\scriptsize{W2V+Mel}};

            \nextgroupplot[
                height=5.0cm,
                width=5.0cm,
                enlargelimits=0.2,
                xlabel={MOS},
                yticklabels={,,},
                title={Group by target feature},
                scatter/classes={%
                        mel={mark=square*,Emerald},
                        ppg={mark=triangle*,Cerulean},
                        apc={mark=*,BurntOrange},
                        cpc={mark=diamond*,RubineRed},
                        w2v={mark=star,draw=Red}
                    },
            ]
            \addplot[
                scatter,mark=*,only marks,
                scatter src=explicit symbolic,
            ]
            table[
                    x=mos,
                    y=sv,
                    meta=target,
                ]{
                    mos   sv     source  target
                    2.90  20.75  mel     ppg
                    2.82  20.25  ppg     ppg
                    3.09  21.75  apc     ppg
                    3.19  10.75  cpc     ppg
                    2.80  4.75   w2v     ppg
                    3.01  92.25  mel     apc
                    2.87  96.00  ppg     apc
                    2.83  92.25  apc     apc
                    3.25  59.00  cpc     apc
                    2.83  11.50  w2v     apc
                    2.55  5.75   mel     w2v
                    3.04  73.75  ppg     w2v
                    3.10  28.50  apc     w2v
                    3.05  35.20  cpc     w2v
                    2.63  7.00   w2v     w2v
                    2.96  98.00  mel     cpc
                    3.19  77.25  ppg     cpc
                    3.14  81.50  apc     cpc
                    3.07  96.75  cpc     cpc
                    2.68  24.75  w2v     cpc
                    2.81  99.50  cpc     mel
                    2.94  69.50  apc     mel
                    2.93  99.50  w2v     mel
                    2.84  92.25  ppg     mel
                    2.74  97.25  mel     mel
                };
            \draw[black] (3.07, 96.75) circle (4.0pt) node[below right, xshift=-2pt, yshift=-0.5pt] {\scriptsize{CPC+CPC}};
            \draw[black] (2.74, 97.25) circle (4.0pt) node[above left, xshift=-1pt, yshift=1pt] {\scriptsize{Mel+Mel}};
            \draw[black] (2.84, 92.25) circle (4.0pt) node[below, yshift=-3pt] {\scriptsize{PPG+Mel}};
            \draw[black] (2.93, 99.50) circle (4.0pt) node[above, yshift=2pt] {\scriptsize{W2V+Mel}};
            \legend{Mel, PPG, APC, CPC, W2V}
        \end{groupplot}
    \end{tikzpicture}
    \caption{
        The overall objective results of MOSNet prediction and Speaker Verification for u2u scenario.
        \emph{W2V} denotes wav2vec 2.0.
    }
    \label{fig:mos_vs_sv}
\end{figure}

As the performance in the u2u scenario is considered more critical than s2s, Fig.~\ref{fig:mos_vs_sv} plots the overall objective results for u2u.
Each model is denoted as A+B, which means taking A as the source feature and B as the target feature.
The points of the two figures are the same, but with different colors.
In the left and right figures, the colors represent the same source feature or target features are used, respectively.
The closer to the upper right means a better model, which achieves better performance in terms of MOSNet prediction and SV results.

Among all models, we can see that the CPC+CPC performs the best, so we select it for the subjective analysis to further explore its ability.
Aside from the CPC+CPC, three models are selected as the baselines in the following subjective analysis.
The first one is Mel+Mel, as the Mel spectrogram has long been used in speech synthesis tasks and showed great performance in VC.
The second one is PPG+Mel, because the PPG is believed to be a better choice in providing content information needed for VC than Mel-spectrogram and performed well in previous PPG-based VC.
The last one is wav2vec 2.0+Mel, which is adopted in FragmentVC and showed excellent performance in VC.

\subsection{Subjective performance analysis}

Subjective evaluation for quality and speaker similarity on the u2u scenario is conducted.
The following four models, including CPC+CPC, Mel+Mel, PPG+Mel, and wav2vec 2.0+Mel are evaluated.

The results are listed in Table~\ref{tab:subjective}, suggesting that CPC+CPC outperforms other baseline models on both quality and speaker similarity, verified that CPC is suitable for providing both content and speaker information needed in VC. 
The converted audio samples are on the demo page~\footnote{https://howard1337.github.io/S2VC/} and the source code will be publicly available~\footnote{https://github.com/howard1337/S2VC}.

\begin{table}
    \setlength{\tabcolsep}{0.2pt} 
    \caption{
        The MOS on unseen-to-unseen conversion.
    }
    \label{tab:subjective}
    \centering
    \begin{tabular}{cccccc}
        \toprule
        \textbf{MOS\,}             &
        {\footnotesize Mel+Mel }   &
        \,{\footnotesize PPG+Mel } &
        \,{\footnotesize W2V+Mel } &
        \,{\footnotesize CPC+CPC}  &
        \,Auth.                                                                                                              \\
        \midrule
        \textbf{Sim.}              & \,2.97$\pm$0.19 & \,3.05$\pm$0.20 & \,3.16$\pm$0.20 & \,3.33$\pm$0.21 & --              \\
        \textbf{Nat.}              & \,2.62$\pm$0.15 & \,3.21$\pm$0.18 & \,2.69$\pm$0.16 & \,3.52$\pm$0.17 & \,4.38$\pm$0.17 \\
        \bottomrule
    \end{tabular}
    
    \justify{
        \emph{A+B}: model with A as source feature and B as target feature. \\
        \emph{W2V}: wav2vec 2.0. \\
        \emph{Auth.}: vocoder-reconstructed authentic utterances.
    }
    \vspace{-1em}
\end{table}

\subsection{Speaker information probing analysis}

As the CPC+CPC model performs well in both objective and subjective evaluation, here we conduct the speaker information probing analysis on the query(Q), key(K), and value(V) for models taking CPC as source feature or as target features.
The speaker classification (SC) task is adopted on the VCTK dataset as the probing task for speaker information.
We randomly sampled 90\% of the VCTK as the training set of SC and the rest 10\% as the development set, and we adopt a linear layer after the extracted feature to classify the speakers.
The training objective for the query feature is to predict the source speaker, while the objective for key and value features is to predict the target speaker.
Both source and target encoder will be used in this analysis, and the source and target feature are ensured to be from different speakers to simulate the inference scenario.

The results are listed in Table~\ref{tab:probing_src}.
The SC accuracy of the query and key features are extremely low, showing that the Instance Normalization and the bottleneck layer can effectively remove the speaker-dependent information.
As for the SC accuracy of the value feature, the models taking CPC as target feature perform better against other models taking PPG, APC, or wav2vec 2.0, showing that CPC can provide rich speaker information needed for VC.


\begin{table}[htb]
    \setlength{\tabcolsep}{2pt} 
    \footnotesize
    \caption{
        Speaker information probing for model taking CPC an source or target feature.
    }
    \label{tab:probing_src}
    \centering
    \begin{tabular}{ccccccccc}
        \toprule

        \multirow{3}{*}{\B{Source}}
                    &
        \multirow{3}{*}{\B{Target}}
                    &
        \multicolumn{1}{c}{\textbf{SC (Query)}} &
        \multicolumn{1}{c}{\textbf{SC (Key)}} &
        \multicolumn{1}{c}{\textbf{SC (Value)}} &
        \\
        \cmidrule(lr){3-3}
        \cmidrule(lr){4-4}
        \cmidrule(lr){5-5}
                & & Dev(\%) & Dev(\%) & Dev(\%) \\
        \midrule
        PPG & \multirow{4}{*}{\textbf{CPC}} & 2.83 & 3.19 & 91.87 \\
        APC & & 2.69 & 3.68 & 91.25 \\
        CPC & & 3.36 & 3.26 & 92.08 \\
        W2V & & 3.31 & 3.77 & 91.36 \\
        \cmidrule(lr){1-5}
        \multirow{4}{*}{\textbf{CPC}} & PPG & 2.74 & 2.09 & 7.15 \\
        & APC & 2.72 & 5.12 & 90.25 \\
        & CPC & 2.36 & 3.26 & 92.08 \\
        & W2v & 2.25 & 2.11 & 70.29 \\
        \bottomrule
    \end{tabular}
    \justify
    \emph{W2V}: wav2vec 2.0.
    \vspace{-1em}
\end{table}

\subsection{Ablation analysis}

We conduct the ablation analysis on the CPC+CPC model.
The SOTA approach FragmentVC~\cite{fragmentvc} is considered as a baseline here.

The ablation results are listed in Table~\ref{tab:ablation}.
It suggests that CPC+CPC outperforms FragmentVC on all metrics considered.
Rows (c) (d) (e) (f) (g) are respectively for removing the self-attention layer, removing the bottleneck layer, removing the instanceNorm layer, removing both bottleneck and instanceNorm layer, and removing cross attention, namely the decoder directly take the output (Q) of the source encoder.
The results verified that all of them are essential for the framework.

\begin{table}[htb]
    \setlength{\tabcolsep}{2pt} 
    \caption{
        Ablation study on the self-attention pooling, bottleneck layer, Instance Normalization and cross attention.
    }
    \label{tab:ablation}
    \centering
    \begin{tabular}{ccccccc}
        \toprule
        \multirow{3}{*}{\textbf{Models}}
                                       &
        \multicolumn{2}{c}{\textbf{MOSNet}}
                                       &
        \multicolumn{2}{c}{\textbf{SV}}
        \\
        \cmidrule(lr){2-3}
        \cmidrule(lr){4-5}
                                       & S2S  & U2U  & S2S(\%) & U2U(\%) \\
        \midrule
        (a) *FragmentVC~\cite{fragmentvc}                & 3.14 & 3.00 & 89.00   & 91.75   \\
        (b) *Proposed                  & 3.36 & 3.07 & 97.75   & 96.75   \\
        (c) *-SAP                      & 3.21 & 2.78 & 97.00   & 95.00   \\
        (d) *-Bottleneck               & 2.90 & 2.49 & 88.00   & 93.25   \\
        (e) *-InstanceNorm             & 3.26 & 2.77 & 98.75   & 95.25   \\
        (f) *-Bottleneck, InstanceNorm & 2.91 & 2.48 & 88.5    & 95.00   \\
        (g) *-Cross attention          & 3.55 & 3.22 & 31.25   & 25.75   \\
        \bottomrule
    \end{tabular}
    \justify
    \emph{FragmentVC} here uses the officially released checkpoint from  https://github.com/yistLin/FragmentVC. \\
\end{table}

\section{Conclusion}
\label{sec:conclusion}

We investigated several SSL representations to improve VC. We found that the model taking CPC as both source and target features outperform the baseline models on both subjective and objective evaluation, including a strong baseline model using PPG as source feature and Mel spectrogram as target feature.
The results suggest that SSL representation CPC is suitable for providing both content and speaker information needed in VC.
Furthermore, the ablation analysis showed that the proposed framework achieves comparable or even better performance than the SOTA approach FragmentVC~\cite{fragmentvc} on objective evaluation.
What will happen if we concatenate several different features like PPG and CPC as source feature and with other combinations of representations as target feature are yet to be investigated. 
We believe that different representations complement each other and provide richer information for both content and speaker information to gain further improvement.

\bibliographystyle{IEEEtran}

\bibliography{main}

\end{document}